\documentclass[aps,twocolumn,superscriptaddress,floatfix,preprintnumbers,longbibliography]{revtex4-1}
\usepackage[utf8]{inputenc}  
\usepackage{amsmath}
\usepackage{amssymb} 
\usepackage{multirow}
\usepackage{enumerate}
\usepackage{amsmath}  
\usepackage{tikz}
\usepackage{tipa}
\usepackage{CJKutf8}
\usepackage{feynmf}
\usepackage{slashed}
\usepackage{braket}
\usepackage[toc,page]{appendix}
\usepackage{url}
\usepackage{natbib}
\usepackage{graphicx}
\usepackage[pdftex,bookmarks,linktocpage,pdfpagelabels,plainpages=false,hyperfigures,linkcolor=blue,citecolor=blue]{hyperref} 
\hypersetup{colorlinks=true}
\usepackage{mathrsfs,amssymb}  
\usepackage{cancel}
\usepackage[normalem]{ulem}
\usepackage{array}
\usepackage{booktabs}
\usepackage{verbatim}

\begin{document}
\title{A Quantum Computing Approach to Track Reconstruction in Strip-Type Detectors}

\author{Seungyeob Jwa}
\email{thoth@snu.ac.kr}
\affiliation{Department of Physics and Astronomy, Seoul National University, Seoul 08826, Republic of Korea}

\author{Hyunyong Kim}
\email{hyunyong@cern.ch}
\thanks{Corresponding author}
\affiliation{Department of Physics and Astronomy, Seoul National University, Seoul 08826, Republic of Korea}

\author{Jangho Kim}
\email{kim.jangho1120@gmail.com}
\thanks{Corresponding author}
\affiliation{Department of Physics and Astronomy, Seoul National University, Seoul 08826, Republic of Korea}

\author{Minseok Oh}
\email{brianohmin@snu.ac.kr}
\affiliation{Department of Physics and Astronomy, Seoul National University, Seoul 08826, Republic of Korea}

\begin{abstract} 
This study investigates the use of quantum annealing for particle track
reconstruction in strip-type gaseous detectors. In such detectors, ghost
hits and multiple hit combinations can turn pattern recognition into a
combinatorial optimization problem. We formulate two reconstruction
subproblems as quadratic unconstrained binary optimization problems. The first
subproblem selects detector hits associated with a single photon track inside
a localized candidate region. The second subproblem selects cluster triplets
from different detector layers so that multiple track candidates can be
handled within a single quantum processing unit submission.

The proposed formulations are tested using simulated DAMSA detector events.
For the single track hit selection task, the QPU based reconstruction gives
position and angular resolutions close to those obtained with a Kalman based
reconstruction. In the simultaneous association task, valid cluster triplets are first extracted from the QPU samples and then connected using an association rule based on graph connectivity to construct track candidates.
The DAMSA event topology studied here has low pileup and is dominated by the two photon signal from ALP decay. In this setting, the results show that the QUBO formulations can reproduce local reconstruction decisions. This provides a practical basis for further studies of reconstruction methods that combine quantum and classical computing in more complex tracking environments.
\end{abstract}

\maketitle

\section{Introduction} 
Although the Standard Model (SM) successfully describes the fundamental particles and their interactions, several important questions remain unresolved, such as the origin of neutrino masses and the nature of dark matter. Many theoretical explanations introduce light bosonic mediators that couple very weakly to SM particles and may connect the visible sector to a dark sector. In particular, MeV-scale mediators are well motivated in portal scenarios, making them important targets for accelerator-based dark-sector searches.

Fixed-target and beam-dump experiments are particularly suitable for probing such light mediators due to their high intensity beams. However, recent studies have shown that these experiments encounter an intrinsic limitation known as the beam-dump ``ceiling,''~\cite{Dutta:2023abe} beyond which improvements in sensitivity to fast-decaying mediators become increasingly difficult. An approach to overcome this limitation is to place the detector very close to the beam target, enabling sensitivity to the prompt-decay region. In such short-baseline configurations, the target length must be reduced to optimize sensitivity to short-lived particles. However, a thinner target and a short detector distance lead to significant leakage of secondary particles, producing a high pileup and high radiation environment near the detector~\cite{Kim:2024vxg,DAMSA:2026fiw}.

Accurate tracking in this environment is therefore essential for signal identification. Strip-type gaseous detectors, such as the micro-Resistive WELL ($\mu$RWell)~\cite{Bencivenni:2014exa}, provide a promising and cost-efficient solution, offering good spatial resolution, radiation tolerance, and scalability for compact detector systems. Nevertheless, strip-based detectors suffer from intrinsic ambiguities known as ghost hits~\cite{Abusleme:2015yja}, which become particularly problematic in high multiplicity environments with significant pileup. These ambiguities make track reconstruction a challenging combinatorial problem.

In this work, we investigate the application of quantum computing to track reconstruction in strip-type detectors. In particular, we explore a quantum annealing approach to solve the optimization problem associated with track finding and fitting in the presence of ghost hits and multiple hit combinations. Our results demonstrate that quantum annealing provides an efficient method for handling the complex combinatorial structure of track reconstruction in high occupancy environments relevant to short-baseline dark-sector experiments.

\section{Quantum Computing}
Quantum computing exploits quantum mechanical phenomena such as superposition, entanglement, and quantum tunneling to perform computations that can differ fundamentally from classical approaches. While gate based quantum computers aim to implement universal quantum circuits, an alternative paradigm known as quantum annealing is specifically designed to solve optimization problems.

Quantum annealing is particularly well suited for combinatorial optimization tasks, where the objective is to find the configuration of variables that minimizes a given cost function. In this approach, the system evolves according to a time-dependent Hamiltonian

\begin{equation}
H(t) = A(t) H_{\mathrm{init}} + B(t) H_{\mathrm{problem}}\,,
\end{equation}
where $H_{\mathrm{init}}$ is the transverse-field Hamiltonian
\begin{equation}
H_{\mathrm{init}} = - \sum_i \sigma_i^{x}\,,
\end{equation}
whose ground state is an equal superposition of all computational basis states. 
The transverse field initializes each qubit in a superposition of $\ket{0}$ and 
$\ket{1}$, enabling the system to explore a large configuration space through 
quantum fluctuations. During the annealing process, the strength of the transverse 
field $A(t)$ is gradually reduced while $B(t)$ increases, guiding the system toward the ground state of the optimization problem.

A commonly used formulation for optimization problems is the Quadratic Unconstrained Binary Optimization (QUBO) representation,
\begin{equation}
E(q) = q^{T} Q q \,,
\end{equation}
where $q$ is a solution vector with binary components $q_i \in \{0,1\}$, and $Q$ is a matrix that encodes the interactions between variables. The objective is to find the binary configuration that minimizes $E(q)$.

\subsection{D-Wave Quantum Annealer and Hardware Embedding}
The D-Wave quantum annealer implements an Ising Hamiltonian on a sparse graph of superconducting flux qubits and programmable couplers. The hardware connectivity determines which couplings $J_{ij}$ can be implemented directly on the device. Earlier D-Wave processors used the Chimera topology, while the Advantage platform is based on the Pegasus topology. The more recent Advantage2 platform employs the Zephyr topology, which increases the native qubit connectivity relative to Pegasus and provides a nominal qubit degree of 20~\cite{Boothby2021Zephyr,McGeoch2022Advantage2}.

Because the hardware graph is sparse, a QUBO or Ising problem whose interaction graph is not directly compatible with the device must be mapped onto the physical qubits by minor embedding~\cite{Choi2011MinorEmbedding}. In this procedure, a logical binary variable is represented by a connected chain of physical qubits with ferromagnetic couplings, so that the chain acts as a single effective variable. This mapping allows interaction graphs that are not native to the hardware to be implemented on the annealer. However, it increases the number of required physical qubits and can modify the effective energy scale of the embedded problem.

In this work, the QUBO generated from the track reconstruction problem is embedded onto the D-Wave hardware graph before annealing. The resulting number of physical qubits and chain lengths are therefore treated as quantities that depend on the implementation and are reported together with the sampling results.

\subsection{Optimization Formulation for Track Reconstruction}
Track reconstruction in strip-type detectors naturally leads to a combinatorial optimization problem. In high-multiplicity environments, multiple hit combinations and ghost hits generate a large number of possible track candidates. The challenge is to select a consistent subset of hit combinations that best describe the underlying particle trajectories.

This task can be formulated as a binary optimization problem by introducing binary variables that represent candidate hit associations. A cost function can then be constructed to encode track-fitting quality, geometrical consistency, and detector constraints. By expressing this cost function in the QUBO form, $E(x) = x^T Q x$, the track reconstruction problem becomes directly compatible with the quantum annealing framework.

In the following section, we describe the specific QUBO formulation used in this work to encode the track reconstruction problem for strip-type detectors and its implementation on the D-Wave quantum annealer.

\section{Benchmark Case}
\subsection{Experimental Configuration}
\begin{figure}[ht]
    \centering
    \includegraphics[width=\linewidth]{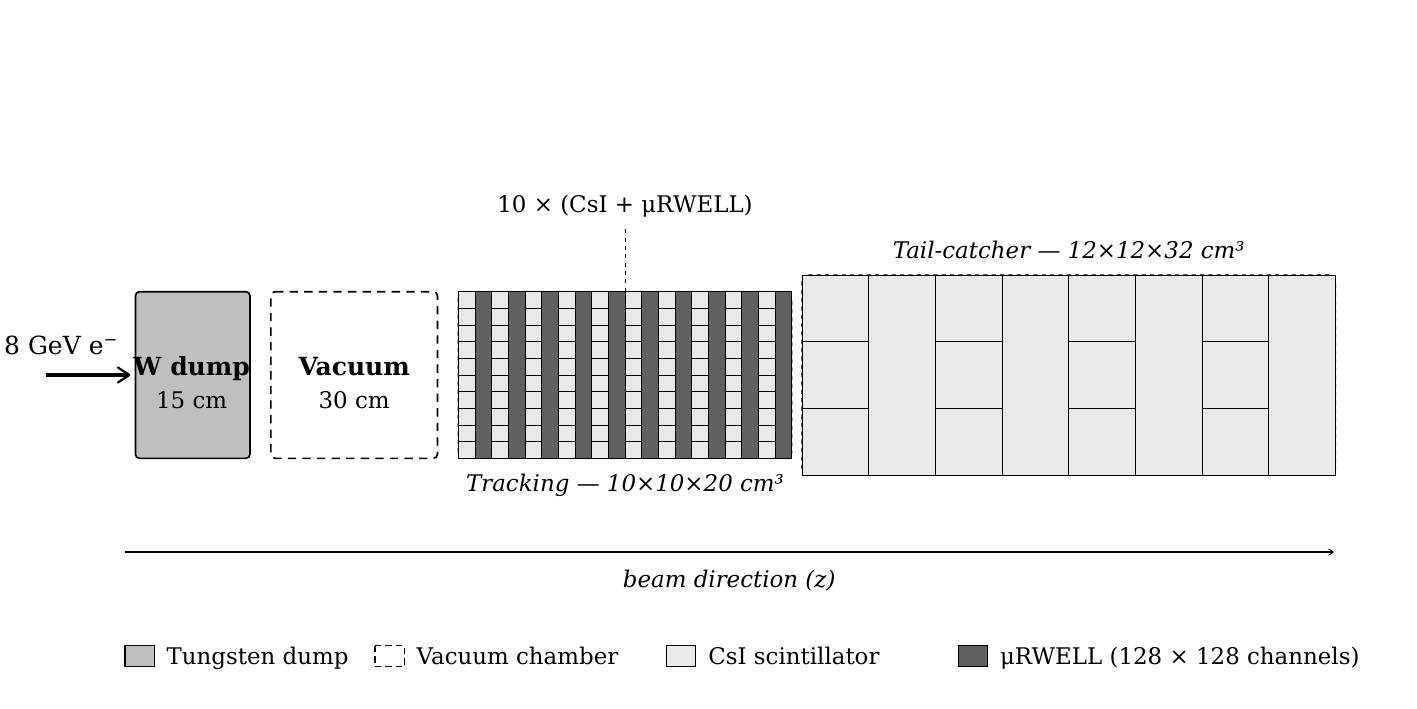}
    \caption{Schematic side view of the compact beam dump experiment. An 8GeV electron beam is stopped in a 15cm tungsten dump, followed by a 30cm vacuum chamber. The ECal comprises a $10\times10\times20 \mathrm{cm}^{3}$ tracking section of ten interleaved 1cm CsI scintillator and 1cm $\mu$RWELL layers, with odd-numbered $\mu$RWELL planes rotated by $45^{\circ}$ to form a $u$-$v$ stereo readout, and a $12\times12\times32\mathrm{cm}^{3}$ tail-catcher of four crossed $x$-$y$ CsI-bar module pairs. The beam propagates along the $z$ direction.}
    \label{fig:detector}
\end{figure}

The experiment is a compact, fixed-target beam dump setup driven by an 8GeV electron beam incident on a tungsten dump of 15cm length, in which the primary beam is stopped and signal candidates are produced. Downstream of the dump, a 30cm long vacuum chamber suppresses interactions of secondary particles before they reach the detector, after which the produced particles enter a newly designed electromagnetic calorimeter (ECal). The ECal consists of two sections: a high-granularity tracking section of $10\times10\times20\mathrm{cm}^{3}$ followed by a tail-catcher section of $12\times12\times32\mathrm{cm}^{3}$, giving a total length of 52~cm along the beam axis. The overall layout is shown in Fig.~\ref{fig:detector}.

The ECal adopts a total-absorption sampling-tracking hybrid concept that combines a fully active CsI medium with high granularity tracking layers, enabling a precise measurement of the photon energy while simultaneously resolving the transverse shower profile. The tracking section is built from ten identical CsI $\mu$RWELL layer pairs stacked along the beam direction, each pair consisting of a 1cm thick CsI scintillator layer followed by a 1cm thick $\mu$RWELL detector, for a total length of 20~cm. Each CsI layer is a $10\times10$ array of $1\times1\times1\mathrm{cm}^{3}$ cubic crystals, providing a finely segmented active medium for electromagnetic shower sampling, while each $\mu$RWELL detector covers $10\times10\mathrm{cm}^{2}$ and is segmented into $128\times128$ readout channels, recording the transverse position of charged shower particles with sub-millimeter precision. To suppress ambiguities (ghost hits) in multiparticle events, the $\mu$RWELL detectors in odd-numbered layers adopt a $u$--$v$ strip orientation rotated by $45^{\circ}$ relative to the regular $x$--$y$ configuration of the even-numbered layers, thereby reducing ghost hits. The ECal is completed by a tail-catcher section composed of four identical $x$-$y$ module pairs stacked along the beam direction; each module pair consists of a first layer of three $4\times4\times12\mathrm{cm}^{3}$ CsI bars oriented along $x$ and stacked in $y$, followed by a second layer of three $4\times4\times12\mathrm{cm}^{3}$ CsI bars oriented along $y$. This crossed layer geometry provides two dimensional containment of the residual shower energy beyond the tracking stack. The total absorption CsI volume ensures full containment of electromagnetic showers for an accurate photon energy reconstruction, while the interleaved $\mu$RWELL layers deliver the high spatial granularity required to discriminate signal topologies from leakage and pileup backgrounds.

\subsection{Geant4 Simulation}

\textsc{Geant4}~\cite{GEANT4:2002zbu} is a simulation toolkit for particle
and matter interactions. The beam-dump simulation employs the
\texttt{FTFP\_BERT} reference physics list, which provides a consistent
description of the electromagnetic and hadronic processes relevant to this
analysis. The signal ALP production is implemented using the DMG4 package,
a \textsc{Geant4}-compatible toolkit for dark-sector particle production in
fixed-target experiments~\cite{Bondi:2021DMG4}. In the benchmark sample used
in this study, the ALP mass and photon coupling are set to
\begin{equation}
m_a = 100~\mathrm{MeV}, \qquad
g_{a\gamma\gamma} = 10^{-4}~\mathrm{GeV}^{-1}.
\end{equation}
DMG4 is used to generate ALPs produced in the tungsten dump and to handle the
visible decay mode $a \rightarrow \gamma\gamma$. The resulting photons are then
propagated through the full detector geometry with \textsc{Geant4}.

Both signal events and the associated leakage particles emerging from the
tungsten dump are propagated through the full detector geometry, and
beam-induced pile-up is incorporated by overlaying multiple primary
interactions consistent with the nominal beam intensity, so that the occupancy
and the underlying activity in each layer are reproduced realistically.
\textsc{Geant4} plays a central role in modeling the conversion of signal
photons into $e^{+}e^{-}$ pairs, the subsequent interactions of electrons and
positrons with the detector material, and the resulting development of the
electromagnetic shower. The shower evolution is propagated through the active
volume to produce simulation hits in the $\mu$RWELL strips, yielding a
faithful representation of the detector response that is subsequently passed
to the digitization and reconstruction chain.

The simulation shows that approximately 99\% of signal photons convert into $e^{+}e^{-}$ pairs within the first 10~cm of CsI, corresponding to the cumulative active depth of the tracking section. After conversion, the electron and positron undergo multiple Coulomb scattering as they traverse the subsequent CsI and $\mu$RWELL layers, which progressively degrades the correlation between their measured trajectories and the original photon direction. To preserve sensitivity to the parent photon kinematics, only the first three to five $\mu$RWELL layers downstream of the conversion vertex are used for direction reconstruction of the $e^{+}e^{-}$ pair, rather than full track fitting over the entire stack. This early-layer approach exploits the high spatial granularity of the $\mu$RWELL readout before multiple scattering dominates, providing an accurate estimate of the incoming photon direction, while the remaining layers and the tail-catcher are reserved for energy containment and shower-shape discrimination.

\section{Track Reconstruction Algorithm}

In this work, two QUBO formulations are studied for detector reconstruction.
The first formulation performs single track hit selection inside a
pre-identified photon cluster candidate. The second formulation performs simultaneous association of cluster candidates from different detector layers by first selecting cluster triplets using the QUBO formulation and then connecting them into track candidates.
These two formulations are not used as consecutive stages of a single
reconstruction chain.  Instead, they are separate demonstrations of how QUBO
optimization can be applied to different reconstruction subproblems.

Both formulations operate on detector objects reconstructed from the simulated
strip and calorimeter response.  They are designed to address combinatorial
ambiguities that arise after detector digitization, rather than to model the
underlying particle interactions themselves.  The use of QUBO and quantum
annealing inspired optimization for charged particle pattern recognition has
also been explored in previous tracking studies
\cite{Bapst2019QuantumAnnealingTracking,Zlokapa2019QuantumTracking}.

\subsection{QUBO Formulation for Single Track Hit Selection}

The first QUBO formulation is designed to select detector hits inside a
localized photon cluster candidate.  The input consists of hit candidates in
three detector layers, where the candidates include both true detector
responses and combinatorial ghost hit candidates.  The goal is to choose one
geometrically consistent hit from each layer.

For a three layer candidate set, two layers are used to define segment
variables, while the remaining layer is used to define hit selection
variables.  In the implementation, the two segment layers are chosen as the
two layers with the smallest candidate multiplicities.  This reduces the
number of segment variables.  Let these two layers be denoted by $a$ and $b$,
and let the remaining layer be denoted by $c$.

A binary segment variable $s_{ij}$ is assigned to the hit pair connecting hit
$i$ in layer $a$ and hit $j$ in layer $b$.  A binary hit variable $h_k$ is
assigned to hit $k$ in layer $c$.  A valid solution must select exactly one
segment and exactly one hit:
\begin{equation}
\sum_{i,j} s_{ij} = 1,
\qquad
\sum_k h_k = 1 .
\end{equation}
These conditions are imposed using quadratic penalty terms, following the
standard QUBO construction of constraints through energy penalties
\cite{Kochenberger2014QUBO}:
\begin{equation}
H_{\mathrm{seg}}
=
A_{\mathrm{seg}}
\left(
\sum_{i,j} s_{ij} - 1
\right)^2 ,
\end{equation}
\begin{equation}
H_{\mathrm{hit}}
=
A_{\mathrm{hit}}
\left(
\sum_k h_k - 1
\right)^2 .
\end{equation}
After dropping constant terms, these constraints are encoded in the QUBO
matrix through diagonal and quadratic terms.

In the present implementation,
\begin{equation}
A_{\mathrm{seg}} = 35.0,
\qquad
A_{\mathrm{hit}} = 35.0 .
\end{equation}
These penalty strengths are chosen to be larger than the maximum straightness
reward, so that invalid configurations with missing or multiple selected
objects are energetically disfavored.  The two constraints are assigned the
same strength because selecting a unique segment and selecting a unique
third-layer hit are equally mandatory requirements of the single track hit
selection problem.

The geometrical consistency of a selected triplet is evaluated by
extrapolating the selected segment $(i,j)$ to the remaining layer $c$:
\begin{equation}
\hat{\mathbf{r}}_{c,ij}
=
\mathbf{r}_{a,i}
+
\frac{z_c-z_a}{z_b-z_a}
\left(
\mathbf{r}_{b,j}
-
\mathbf{r}_{a,i}
\right),
\end{equation}
where
\begin{equation}
\mathbf{r}_{l,i} = (x_{l,i},y_{l,i}) .
\end{equation}
The residual of the triplet $(i,j,k)$ is then defined as
\begin{equation}
d_{ijk}
=
\left|
\mathbf{r}_{c,k}
-
\hat{\mathbf{r}}_{c,ij}
\right| .
\end{equation}
Smaller values of $d_{ijk}$ correspond to more linear hit combinations.

To make the straightness reward comparable over the full candidate set, the
residuals are normalized using the minimum and maximum residual values
computed from all candidate triplets considered in the reconstruction problem:
\begin{equation}
d_{\min}
=
\min_{i,j,k} d_{ijk},
\qquad
d_{\max}
=
\max_{i,j,k} d_{ijk}.
\end{equation}

The normalized residual is
\begin{equation}
\tilde{d}_{ijk}
=
\frac{
d_{ijk} - d_{\min}
}{
\max\left(
d_{\max} - d_{\min},
10^{-12}
\right)
}.
\end{equation}
The straightness reward is defined as
\begin{equation}
R_{ijk}
=
1-\tilde{d}_{ijk}.
\end{equation}
Thus, the most linear triplets receive the largest reward, while geometrically
poor triplets receive little or no reward.  The corresponding QUBO term is
\begin{equation}
H_{\mathrm{line}}
=
-
B_{\mathrm{line}}
\sum_{i,j,k}
R_{ijk}
s_{ij}h_k ,
\end{equation}
with
\begin{equation}
B_{\mathrm{line}} = 10.0 .
\end{equation}
The value of $B_{\mathrm{line}}$ is chosen smaller than the constraint
penalties, so that straightness ranks valid configurations without overriding
the one segment and one hit constraints.

The full single track hit selection QUBO energy is therefore
\begin{align}
H_{\mathrm{single}}
=&\;
A_{\mathrm{seg}}
\left(
\sum_{i,j} s_{ij} - 1
\right)^2
+
A_{\mathrm{hit}}
\left(
\sum_k h_k - 1
\right)^2
\nonumber\\
&-
B_{\mathrm{line}}
\sum_{i,j,k}
\left(
1-\tilde{d}_{ijk}
\right)
s_{ij}h_k .
\end{align}

In the practical implementation, each detector layer may contain more than
four hit candidates.  To keep the QUBO size bounded, the hit candidates in
each layer are divided into chunks containing at most four hits.  A QUBO block
is constructed from one chunk in each of the three layers.

If the candidate multiplicities in the three selected chunks are $N_a$,
$N_b$, and $N_c$, the number of logical variables in one QUBO block is
\begin{equation}
N_{\mathrm{var}}^{\mathrm{single}}
=
N_aN_b + N_c .
\end{equation}
Since each chunk contains at most four hits,
\begin{equation}
N_{\mathrm{var}}^{\mathrm{single}}
\le
4\times4 + 4
=
20 .
\end{equation}
The number of logical quadratic couplings is also bounded.  The exactly one
segment constraint contributes at most
\begin{equation}
{N_aN_b \choose 2}
=
\frac{N_aN_b(N_aN_b-1)}{2}
\end{equation}
quadratic couplings, the exactly one hit constraint contributes at most
\begin{equation}
{N_c \choose 2}
=
\frac{N_c(N_c-1)}{2}
\end{equation}
quadratic couplings, and the straightness reward contributes at most
\begin{equation}
N_aN_bN_c
\end{equation}
segment-hit couplings.  Therefore,
\begin{equation}
N_{\mathrm{coup}}^{\mathrm{single}}
\le
{16 \choose 2}
+
{4 \choose 2}
+
16\times4
=
190 .
\end{equation}

Several independent QUBO blocks are assembled into a single block diagonal
QUBO matrix and submitted together to the quantum annealer.  This is used as
a batching strategy to reduce the number of QPU calls.  The physical
reconstruction unit remains the individual three layer QUBO block.  In the
implementation, at most 50 local blocks are included in one QPU submission.
Thus, the submitted logical problem size is bounded by
\begin{equation}
N_{\mathrm{var}}^{\mathrm{sub}}
\le
50 \times 20
=
1000 ,
\end{equation}
and the logical coupler count is bounded by
\begin{equation}
N_{\mathrm{coup}}^{\mathrm{sub}}
\le
50 \times 190
=
9500 .
\end{equation}
The actual numbers of logical variables and logical couplers are recorded for
each QPU submission.  The logical coupler count corresponds to the number of
nonzero off diagonal entries in the QUBO matrix and is not the same as the
number of physical qubits used on the quantum annealer.  The physical qubit
usage depends on the minor embedding and the resulting chain lengths.

If the number of local blocks exceeds the maximum allowed number of blocks per
submission, the candidate chunks are split and submitted in multiple QPU
calls.  The same normalization constants $d_{\min}^{\mathrm{all}}$ and
$d_{\max}^{\mathrm{all}}$ are used for all split submissions, so that the
straightness rewards remain comparable across the full candidate set.

After sampling, all returned bitstrings are decoded block by block.  A decoded
block solution is accepted as valid only when exactly one segment variable and
exactly one hit variable are selected.  For each block, the valid solution
with the lowest block QUBO energy is retained among all returned samples.  The
retained block solutions are then merged across all submissions, and the final
detector hit candidate is selected according to the lowest block QUBO energy,
with the geometrical residual used only as a tie breaker.

\subsection{QUBO Formulation for Simultaneous Track Candidate Association}

The second QUBO formulation is designed to associate cluster candidates from
different detector layers into track candidates.  The input to this
formulation is a set of cluster candidates obtained after the detector hit
filtering step.  The surviving hits are grouped in the transverse
$(x,y)$ plane, and each group is represented by its detector layer and
cluster center.  This differs from the single track hit selection formulation,
where the input consists of raw detector hit candidates inside a localized
photon cluster region.

The main purpose of this formulation is to test whether several local track
candidates can be selected in the same QPU submission.  In a given three layer
block, the QPU output is a set of cluster triplets.  Multiple cluster
triplets can be reconstructed at once, and many independent three layer
blocks are submitted together as a block diagonal QUBO matrix.  Thus, the QPU
is used not only to choose a single best association, but to solve a set of
cluster association problems simultaneously.

For detector layer $l$, the cluster candidates are denoted by
\begin{equation}
\mathbf{x}_{l,i} = (x_{l,i},y_{l,i},z_l),
\end{equation}
where $i$ labels the cluster candidate in that layer.  All possible three
layer combinations $(a,b,c)$ with $a<b<c$ are considered.  For ten detector
layers, the number of possible three layer blocks is
\begin{equation}
{10 \choose 3} = 120 .
\end{equation}
Each three layer block is treated as the basic reconstruction unit.

For a given block $(a,b,c)$, edge variables are introduced between layers
$a$ and $b$, and between layers $b$ and $c$.  A binary variable
$e^{ab}_{ij}$ is assigned to the association between cluster $i$ in layer
$a$ and cluster $j$ in layer $b$.  Similarly, a binary variable
$e^{bc}_{jk}$ is assigned to the association between cluster $j$ in layer
$b$ and cluster $k$ in layer $c$.  A cluster triplet is formed when the two
selected edges share the same middle layer cluster:
\begin{equation}
(a,i) \rightarrow (b,j) \rightarrow (c,k).
\end{equation}

The first QUBO energy term rewards the selection of edge variables:
\begin{equation}
H_{\mathrm{edge}}
=
-
R_{\mathrm{edge}}
\left(
\sum_{i,j} e^{ab}_{ij}
+
\sum_{j,k} e^{bc}_{jk}
\right).
\end{equation}
In the present implementation,
\begin{equation}
R_{\mathrm{edge}} = 10.0 .
\end{equation}
This reward favors cluster triplet formation over empty edge configurations.

Branching and merging are suppressed using degree penalties.  For the first
layer, multiple outgoing edges from the same cluster are penalized:
\begin{equation}
H_{\mathrm{deg}}^{a}
=
A_{\mathrm{deg}}
\sum_i
\sum_{j_1<j_2}
e^{ab}_{ij_1}e^{ab}_{ij_2}.
\end{equation}
For the last layer, multiple incoming edges to the same cluster are penalized:
\begin{equation}
H_{\mathrm{deg}}^{c}
=
A_{\mathrm{deg}}
\sum_k
\sum_{j_1<j_2}
e^{bc}_{j_1k}e^{bc}_{j_2k}.
\end{equation}
The degree penalty strength is set to
\begin{equation}
A_{\mathrm{deg}} = 30.0 .
\end{equation}
This value is chosen to be larger than the straightness reward, so that
ambiguous branching or merging associations are energetically disfavored
before geometrical straightness is optimized.

The continuity of a triplet through the middle layer is enforced by a flow
constraint.  For each middle layer cluster $j$, the number of selected
incoming edges and the number of selected outgoing edges should be equal:
\begin{equation}
\sum_i e^{ab}_{ij}
=
\sum_k e^{bc}_{jk}.
\end{equation}
This condition is encoded as
\begin{equation}
H_{\mathrm{flow}}
=
F_{\mathrm{flow}}
\sum_j
\left(
\sum_i e^{ab}_{ij}
-
\sum_k e^{bc}_{jk}
\right)^2 .
\end{equation}
In the present implementation,
\begin{equation}
F_{\mathrm{flow}} = 20.0 .
\end{equation}
The flow penalty is also larger than the straightness reward, because a
selected incoming edge without a matching outgoing edge does not correspond to
a valid cluster triplet.

The geometrical quality of a selected triplet is evaluated using the change
in direction between the two selected segments.  For a triplet $(i,j,k)$, the
normalized direction vectors are
\begin{equation}
\mathbf{u}_{ij}
=
\frac{
\mathbf{x}_{b,j}-\mathbf{x}_{a,i}
}{
\left|
\mathbf{x}_{b,j}-\mathbf{x}_{a,i}
\right|
},
\qquad
\mathbf{u}_{jk}
=
\frac{
\mathbf{x}_{c,k}-\mathbf{x}_{b,j}
}{
\left|
\mathbf{x}_{c,k}-\mathbf{x}_{b,j}
\right|
}.
\end{equation}
The bending measure is defined as
\begin{equation}
b_{ijk}
=
\left|
\mathbf{u}_{jk}
-
\mathbf{u}_{ij}
\right|^2 .
\end{equation}
Smaller values of $b_{ijk}$ correspond to straighter three dimensional
cluster combinations.

To make the bending reward comparable among the candidate triplets included in a single QPU submission, the bending values are normalized using the minimum and maximum bending values in that submission:
\begin{equation}
b_{\min}
=
\min_{i,j,k} b_{ijk},
\qquad
b_{\max}
=
\max_{i,j,k} b_{ijk}.
\end{equation}
The normalized bending value is
\begin{equation}
\tilde{b}_{ijk}
=
\frac{
b_{ijk}-b_{\min}
}{
\max\left(
b_{\max}-b_{\min},
10^{-12}
\right)
}.
\end{equation}
The corresponding straightness reward is
\begin{equation}
H_{\mathrm{bend}}
=
-
B_{\mathrm{bend}}
\sum_{i,j,k}
\left(
1-\tilde{b}_{ijk}
\right)
e^{ab}_{ij}e^{bc}_{jk}.
\end{equation}
In the present implementation,
\begin{equation}
B_{\mathrm{bend}} = 5.0 .
\end{equation}
The value of $B_{\mathrm{bend}}$ is smaller than the degree and flow
penalties, so that the bending term ranks valid triplet candidates without
overriding the topology constraints.

Combining the edge reward, degree constraints, flow constraint, and
straightness reward, the block QUBO energy is
\begin{align}
H_{\mathrm{block}}
=&\;
-R_{\mathrm{edge}}
\left(
\sum_{i,j} e^{ab}_{ij}
+
\sum_{j,k} e^{bc}_{jk}
\right)
\nonumber\\
&+
A_{\mathrm{deg}}
\sum_i
\sum_{j<j'}
e^{ab}_{ij} e^{ab}_{ij'}
\nonumber\\
&+
A_{\mathrm{deg}}
\sum_k
\sum_{j<j'}
e^{bc}_{jk} e^{bc}_{j'k}
\nonumber\\
&+
F_{\mathrm{flow}}
\sum_j
\left(
\sum_i e^{ab}_{ij}
-
\sum_k e^{bc}_{jk}
\right)^2
\nonumber\\
&-
B_{\mathrm{bend}}
\sum_{i,j,k}
\left(
1-\tilde{b}_{ijk}
\right)
e^{ab}_{ij} e^{bc}_{jk}.
\end{align}

If the cluster multiplicities in the three layers are $N_a$, $N_b$, and
$N_c$, the number of logical variables in one three layer block is
\begin{equation}
N_{\mathrm{var}}^{\mathrm{triplet}}
=
N_aN_b + N_bN_c .
\end{equation}
The first term counts the edge variables between layers $a$ and $b$, and the
second term counts the edge variables between layers $b$ and $c$.

The number of logical quadratic couplings is bounded by the degree, flow, and
bending terms.  The first layer degree penalty contributes
\begin{equation}
N_a {N_b \choose 2}
=
N_a
\frac{N_b(N_b-1)}{2}
\end{equation}
couplings, while the last layer degree penalty contributes
\begin{equation}
N_c {N_b \choose 2}
=
N_c
\frac{N_b(N_b-1)}{2}
\end{equation}
couplings.  The flow term contributes, for each middle layer cluster,
couplings among the incoming edges, among the outgoing edges, and between
incoming and outgoing edges:
\begin{equation}
N_b
\left[
{N_a \choose 2}
+
{N_c \choose 2}
+
N_aN_c
\right].
\end{equation}
Finally, the bending reward contributes one coupling for each triplet:
\begin{equation}
N_aN_bN_c .
\end{equation}

Several three layer blocks are assembled into a single block diagonal QUBO
matrix and submitted together to the quantum annealer.  This batching strategy
reduces the number of QPU calls while preserving the individual three layer
block as the physical reconstruction unit.  In the implementation, at most 50
three layer blocks are included in one QPU submission.  If more than 50 blocks
are present, the blocks are split as evenly as possible among multiple QPU
submissions.  For ten detector layers, the 120 three layer combinations are
therefore split into three submissions of 40 blocks each.

After sampling, all returned bitstrings are decoded block by block.  For a
block $(a,b,c)$, the expected number of selected cluster triplets is
\begin{equation}
M_{abc}
=
\min(N_a,N_b,N_c).
\end{equation}
A decoded block solution is accepted as valid only when it contains exactly
$M_{abc}$ complete cluster triplets through the middle layer.  The selected
edges must not contain branching, merging, or dangling edges.  Thus, every
accepted block solution can be written as
\begin{equation}
\mathcal{T}_{abc}
=
\left\{
\tau_{abc}^{(1)}, \tau_{abc}^{(2)}, \ldots,
\tau_{abc}^{(M_{abc})}
\right\},
\end{equation}
where each triplet has the form
\begin{equation}
\tau_{abc}^{(m)}
:
(a,i) \rightarrow (b,j) \rightarrow (c,k).
\end{equation}
Each triplet spans all three layers of the block.  Two layer fragments are not
accepted as reconstructed triplets in this formulation.

This construction allows more than one local track candidate to be decoded in
the same QPU submission.  For the ALP signal topology, in which the ALP decays
into two photons, a valid block solution can contain two simultaneous photon
triplets,
\begin{equation}
\mathcal{T}_{abc}^{\gamma\gamma}
=
\left\{
\tau_{abc}^{(1)}, \tau_{abc}^{(2)}
\right\},
\end{equation}
with each $\tau_{abc}^{(r)}$ corresponding to one three layer photon
candidate.  For each block, all valid decoded solutions from all returned
samples are inspected, and the valid solution with the lowest block QUBO
energy is retained.  Each triplet in the retained set $\mathcal{T}_{abc}$ is
then stored as an individual seed for the subsequent graph based association
step.

The seed triplets are then assembled into longer track candidates using a
graph overlap rule.  Each seed triplet is treated as a graph vertex, and an
undirected edge is placed between two seed triplets if they share at least two
identical layer-cluster assignments.  That is, two seeds are connected when
\begin{equation}
\left|
S_p \cap S_q
\right|
\geq 2 ,
\end{equation}
where
\begin{equation}
S_p =
\{(l_1,c_1),(l_2,c_2),(l_3,c_3)\}
\end{equation}
is the set of layer-cluster nodes contained in seed triplet $p$.

The final track candidates are obtained as the connected components of this
seed overlap graph.  In the implementation, the connected components are
found using a standard disjoint set procedure~\cite{Tarjan1975SetUnion}.  For
each connected component, the union of all layer-cluster assignments in that
component defines the reconstructed track candidate:
\begin{equation}
T_\alpha
=
\bigcup_{p \in C_\alpha} S_p ,
\end{equation}
where $C_\alpha$ denotes one connected component of the graph.  This
postprocessing step does not perform an additional line fit, Kalman update, or
residual based optimization.  It only combines QPU selected triplets using
exact shared-node consistency.

This construction separates the role of the quantum annealer from the
subsequent graph assembly step.  The quantum annealer is responsible for selecting
geometrically consistent local cluster triplets, including multiple triplets
within the same block and multiple blocks within the same QPU submission.  The
graph step only links overlapping local triplets into longer track candidates.
Therefore, the performance of the method is evaluated both at the cluster
triplet level and at the graph assembled track candidate level.

\section{Results}

\subsection{Performance of the quantum annealer Single Track Hit Selection}

The performance of the QPU single track hit selection method described in the
previous section is evaluated by comparing the reconstructed track parameters
with those obtained from a Kalman filter reconstruction.  Kalman filtering is
widely used for track and vertex reconstruction in particle detectors
\cite{Fruhwirth1987Kalman}.  In this study, both methods are applied to the
same single photon reconstruction sample.  The Kalman reconstruction is used
as a conventional baseline, while the QUBO method performs the discrete block
hit selection before the final track parameters are reconstructed.

Figures~\ref{fig:res_r_compare} and~\ref{fig:res_theta_compare} show the
resulting position and angular residual distributions.  The Kalman sample
contains 1133 reconstructed events, while the QPU sample contains 1128
reconstructed events after the corresponding reconstruction requirements.

\subsubsection{Residual Definitions}

The position residual is defined at the production plane.  The reconstructed
track is extrapolated back to the $z=0$ plane, and the transverse distance
between the extrapolated reconstructed position and the truth position is
calculated as
\begin{align}
\Delta R
=
\bigg[
&\left(x_{\mathrm{reco}}(z=0)-x_{\mathrm{truth}}(z=0)\right)^2
\nonumber\\
&+
\left(y_{\mathrm{reco}}(z=0)-y_{\mathrm{truth}}(z=0)\right)^2
\bigg]^{1/2}.
\end{align}

The angular residual is defined as the opening angle between the reconstructed
track direction and the truth direction,
\begin{equation}
\Delta \theta
=
\cos^{-1}
\left(
\hat{\mathbf{u}}_{\mathrm{reco}}
\cdot
\hat{\mathbf{u}}_{\mathrm{truth}}
\right),
\end{equation}
where $\hat{\mathbf{u}}_{\mathrm{reco}}$ and
$\hat{\mathbf{u}}_{\mathrm{truth}}$ are unit direction vectors.

The displayed histogram ranges correspond to the central 95\% of each
distribution in order to emphasize the core reconstruction performance.  All
quoted statistical quantities, however, are evaluated using the full
distributions before applying the plotting range selection.  Since the
residual distributions contain non-Gaussian tails from ambiguous or
incorrectly matched hit combinations, both full distribution quantities and
robust core resolution metrics are reported.

\subsubsection{Position Resolution}

\begin{figure}[htbp]
    \centering
    \includegraphics[width=\linewidth]{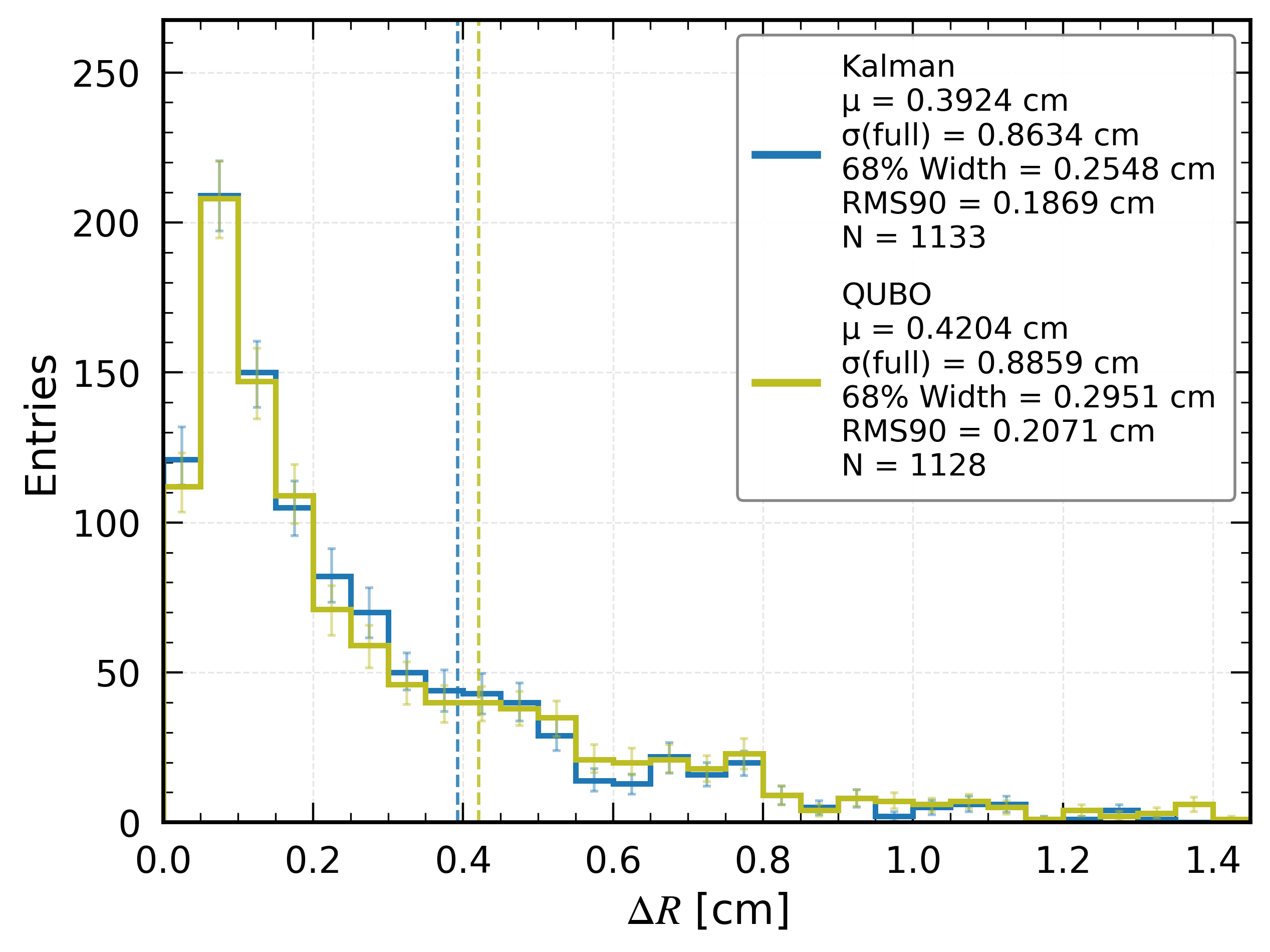}
    \caption{
        Comparison of the position residual distributions for the Kalman and
        QPU reconstruction methods. The reconstructed track is extrapolated to
        the $z=0$ plane, and the residual is computed as the transverse
        distance from the truth position. The dashed vertical lines indicate
        the mean residual values.
    }
    \label{fig:res_r_compare}
\end{figure}

Figure~\ref{fig:res_r_compare} shows the position residual distributions.  The
two methods exhibit very similar overall shapes, indicating that the QUBO
block selection reconstructs track candidates with a performance close to the
Kalman filter approach.

For the full distributions, the Kalman reconstruction gives a mean position
residual of $0.3925~\mathrm{cm}$ with
$\sigma_{\mathrm{full}} = 0.8634~\mathrm{cm}$, while the QPU reconstruction
gives a slightly larger mean residual of $0.4204~\mathrm{cm}$ with
$\sigma_{\mathrm{full}} = 0.8860~\mathrm{cm}$.

The robust core resolution metrics show the same trend.  The Kalman
reconstruction gives a 68\% width of $0.2548~\mathrm{cm}$ and an RMS90 value
of $0.1869~\mathrm{cm}$, while the QPU reconstruction gives a 68\% width of
$0.2951~\mathrm{cm}$ and an RMS90 value of $0.2071~\mathrm{cm}$.  Thus, the
Kalman method provides a slightly narrower core response, but the degradation
observed for the QUBO method remains moderate.

A bin by bin comparison further shows that the two position residual
distributions are statistically similar.  The average bin difference is
$0.630\sigma$, with a maximum local difference of $2.625\sigma$.  The
histogram comparison gives $\chi^2/N_{\mathrm{dof}}=21.38/26$,
corresponding to a reduced $\chi^2$ of 0.822.  The KL divergence is
$D_{\mathrm{KL}}=0.0178$.  These values indicate that the QPU reconstruction
reproduces the overall shape of the Kalman residual distribution, with only
small differences in the core width and tail behavior.

\subsubsection{Angular Resolution}

\begin{figure}[htbp]
    \centering
    \includegraphics[width=\linewidth]{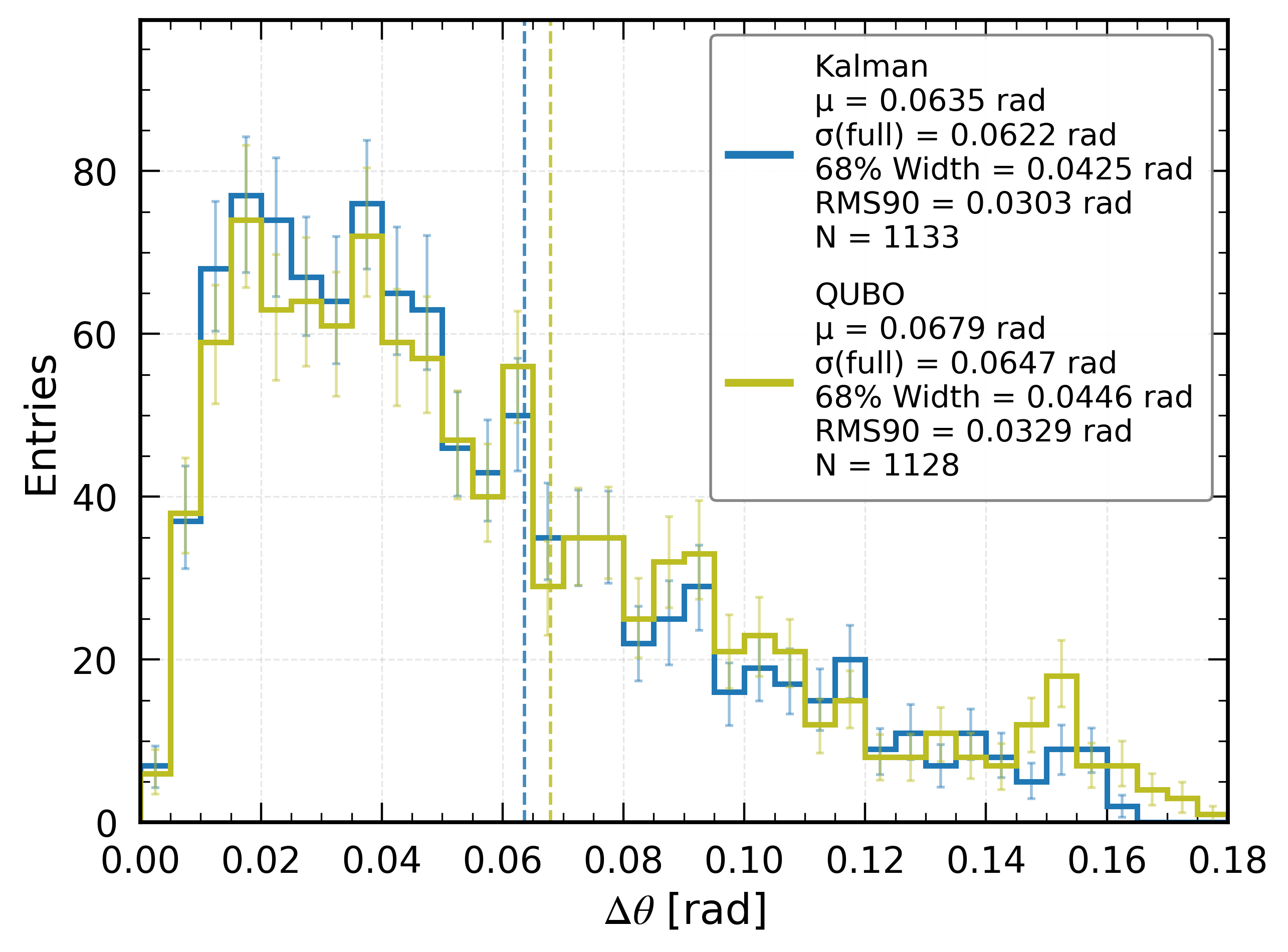}
    \caption{
        Comparison of the angular residual distributions for the Kalman and
        QPU reconstruction methods. The angular residual is defined as the
        opening angle between the reconstructed track direction and the truth
        direction. The dashed vertical lines indicate the mean angular
        residual values.
    }
    \label{fig:res_theta_compare}
\end{figure}

Figure~\ref{fig:res_theta_compare} shows the angular residual distributions.
Compared with the position residual case, the agreement between the two
reconstruction methods is similarly strong.

For the full angular residual distributions, the Kalman reconstruction gives a
mean angular residual of $0.0635~\mathrm{rad}$ with
$\sigma_{\mathrm{full}} = 0.0622~\mathrm{rad}$, while the QPU reconstruction
gives $0.0679~\mathrm{rad}$ with
$\sigma_{\mathrm{full}} = 0.0647~\mathrm{rad}$.  The corresponding 68\%
widths are $0.0425~\mathrm{rad}$ and $0.0446~\mathrm{rad}$ for the Kalman and
QPU methods, respectively.  The RMS90 values are $0.0303~\mathrm{rad}$ for
the Kalman reconstruction and $0.0329~\mathrm{rad}$ for the QPU
reconstruction.

The angular shape comparison also indicates strong consistency between the
two methods.  The average bin difference is $0.622\sigma$, and the maximum
local difference is $1.808\sigma$.  The histogram comparison gives
$\chi^2/N_{\mathrm{dof}}=26.82/33$, corresponding to a reduced $\chi^2$ of
0.813.  The KL divergence is $D_{\mathrm{KL}}=0.0241$.  These results show
that the angular residual distributions are very similar within the
statistical precision of the sample.

\subsubsection{Summary of Resolution Metrics}

\begin{table}[htbp]
    \centering
    \begin{tabular}{c|cc|cc}
        \hline
        \multirow{2}{*}{Metric}
        & \multicolumn{2}{c|}{Position residual [cm]}
        & \multicolumn{2}{c}{Angular residual [rad]} \\
        & Kalman & QPU & Kalman & QPU \\
        \hline
        Mean & 0.3925 & 0.4204 & 0.0635 & 0.0679 \\
        $\sigma_{\mathrm{full}}$ & 0.8634 & 0.8860 & 0.0622 & 0.0647 \\
        68\% width & 0.2548 & 0.2951 & 0.0425 & 0.0446 \\
        RMS90 & 0.1869 & 0.2071 & 0.0303 & 0.0329 \\
        \hline
    \end{tabular}
    \caption{
        Summary of the position and angular residual metrics for the Kalman
        and QPU reconstruction methods. The position residual is evaluated at
        the $z=0$ plane, and the angular residual is evaluated from the angle
        between the reconstructed and truth directions. All values are computed
        using the full residual distributions.
    }
    \label{tab:resolution_summary}
\end{table}

\begin{table}[htbp]
    \centering
    \begin{tabular}{c|cc}
        \hline
        Shape comparison metric & Position residual & Angular residual \\
        \hline
        Average bin difference & $0.630\sigma$ & $0.622\sigma$ \\
        Maximum bin difference & $2.625\sigma$ & $1.808\sigma$ \\
        $\chi^2/N_{\mathrm{dof}}$ & $21.38/26$ & $26.82/33$ \\
        Reduced $\chi^2$ & 0.822 & 0.813 \\
        $D_{\mathrm{KL}}$ & 0.0178 & 0.0241 \\
        \hline
    \end{tabular}
    \caption{
        Shape comparison between the Kalman and QPU residual distributions.
        Here, $N_{\mathrm{dof}}$ denotes the number of degrees of freedom used
        in the bin by bin histogram comparison.
    }
    \label{tab:shape_comparison}
\end{table}

Overall, the Kalman reconstruction gives the best numerical resolution in both
position and angle, as expected from its continuous state update procedure.
Nevertheless, the quantum annealing based hit selection method achieves very similar residual
distributions using a discrete block formulation.  The reduced $\chi^2$ values
below unity and the low KL divergences show that the QUBO method reproduces
the Kalman residual shapes closely.  These results demonstrate that the QUBO
formulation is a viable approach for ghost hit rejection and detector hit
selection in this detector environment.

\subsubsection{QPU Resource Scaling for Single Track Hit Selection}

The QPU embedding resource usage was evaluated using the submitted block
matrix QUBO problems.  Since each QPU call can contain a different number of
independent three layer QUBO blocks, the resource usage is normalized by the
number of blocks in the submitted problem.  The per block logical variable
count and physical qubit usage are defined as
\begin{equation}
N_{\mathrm{logical}}^{\mathrm{block}}
=
\frac{N_{\mathrm{logical}}}{N_{\mathrm{block}}},
\qquad
N_{\mathrm{physical}}^{\mathrm{block}}
=
\frac{N_{\mathrm{physical}}}{N_{\mathrm{block}}}.
\end{equation}
This normalization removes the trivial increase caused by submitting more
local blocks in a single QPU call, and instead measures the embedding cost of
a single local reconstruction unit.

\begin{figure}[htbp]
    \centering
    \includegraphics[width=\linewidth]{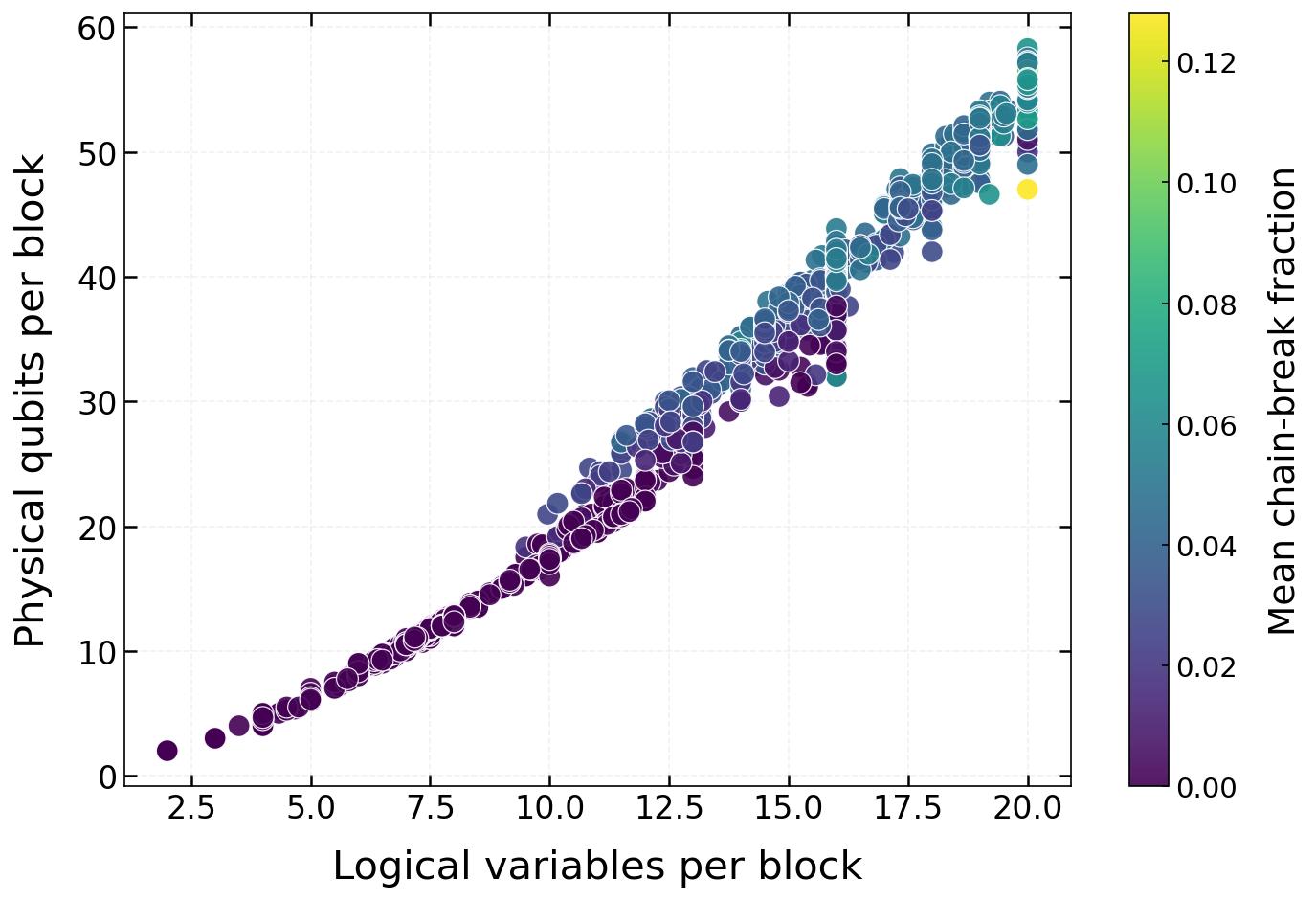}
    \caption{
        QPU embedding resource usage for the single track hit selection QUBO.
        Both axes are normalized by the number of local QUBO blocks in the
        submitted problem. The plot shows the physical qubit cost per block as
        a function of the logical variable count per block.
    }
    \label{fig:qubo_hit_resource_scaling}
\end{figure}

Figure~\ref{fig:qubo_hit_resource_scaling} shows the number of physical
qubits used per block as a function of the number of logical variables per
block.  The approximately monotonic trend indicates that the required
embedding resource grows smoothly with the local QUBO size.  This behavior
shows that the block matrix formulation remains controlled on the quantum
annealer while allowing multiple independent detector hit selection problems
to be submitted together.

\subsubsection{QPU Timing Model and Latency Deconstruction}

To evaluate the operational efficiency of the QPU during the single track hit selection phase, we deconstruct the total latency into three distinct physical components: queue/access overhead ($T_{\mathrm{overhead}}$), embedding and programming overhead ($T_{\mathrm{programming}}$), and quantum annealing/sampling latency ($T_{\mathrm{sampling}}$). The total end-to-end latency per single-photon execution, denoted as $T_{\mathrm{total}}$, is formulated as:
\begin{equation}
\label{eq:t_total}
T_{\mathrm{total}} = T_{\mathrm{overhead}} + T_{\mathrm{programming}} + T_{\mathrm{sampling}}.
\end{equation}
The core quantum sampling process, which represents the quantum state evolution on the superconducting lattice, scales directly with the number of reads $R$ and is physically bounded by the single-sample duration $T_{\mathrm{sample}}$:
\begin{equation}
\label{eq:t_sampling}
T_{\mathrm{sampling}} = S \times R \times T_{\mathrm{sample}},
\end{equation}
where $S$ represents the number of applied spin reversal transforms, and $T_{\mathrm{sample}}$ is the hardware-level timing loop determined by the annealing time ($T_{\mathrm{a}}$), readout time ($T_{\mathrm{r}}$), and thermalization delay ($T_{\mathrm{d}}$):
\begin{equation}
\label{eq:t_sample}
T_{\mathrm{sample}} = T_{\mathrm{a}} + T_{\mathrm{r}} + T_{\mathrm{d}}.
\end{equation}

For this photon-level evaluation, a rapid sampling configuration was deployed to maximize execution throughput across the hit candidates, setting $T_{\mathrm{a}} = 20.0\,\mu\mathrm{s}$, $R = 200$ reads, and $S = 1$ spin reversal transform. Under this physical setup, the measured programming cost per photon event is $T_{\mathrm{programming}} = 44.9\,\mathrm{ms}$. This optimized timing envelope yields an average quantum annealing and sampling latency of $T_{\mathrm{sampling}} = 41.9\,\mathrm{ms}$, resulting in a total mean end-to-end latency of $96.5\,\mathrm{ms}$ per track, as deconstructed in Fig. 5.

\begin{figure}[htbp]
    \centering
    \includegraphics[width=0.75\linewidth]{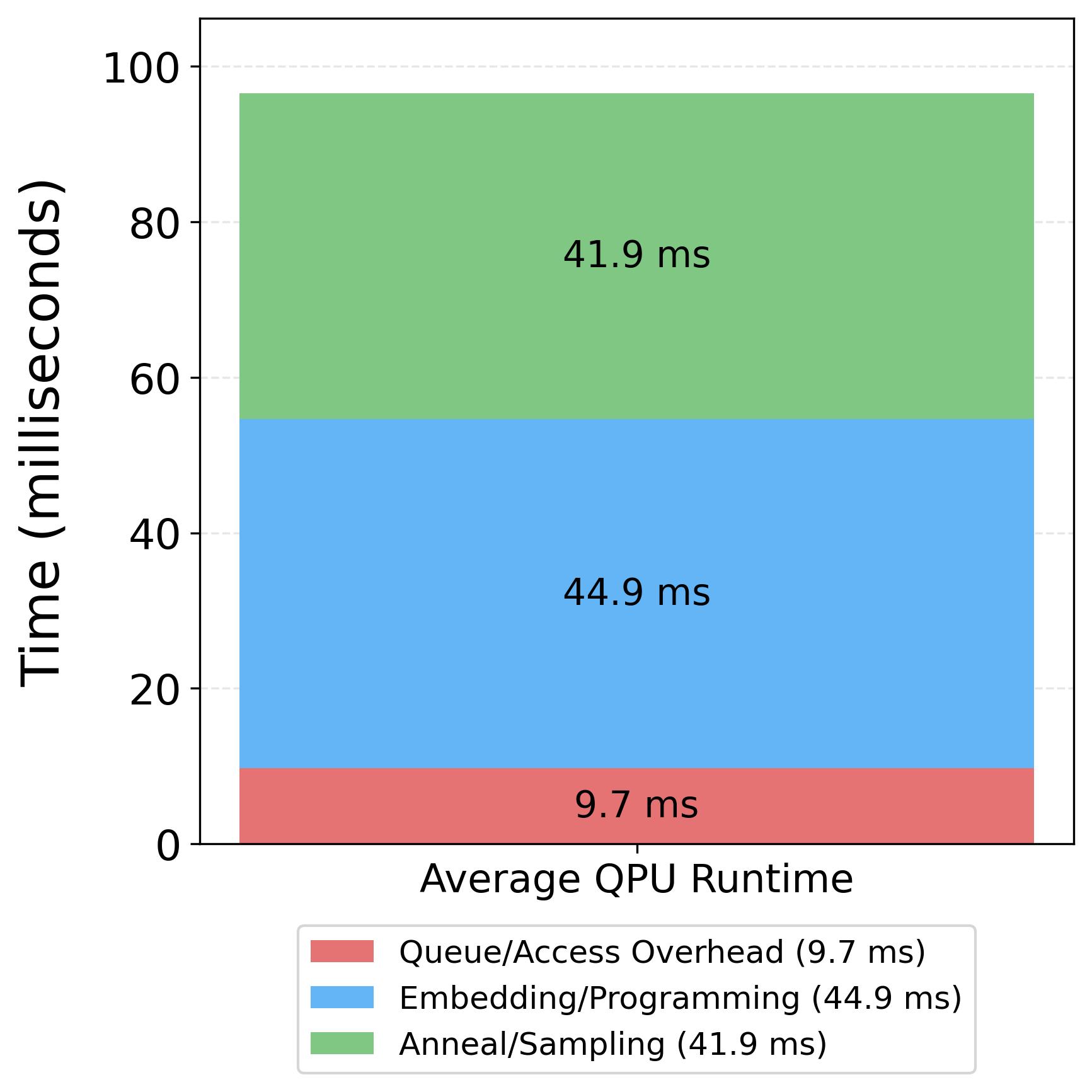}
    \caption{
        Deconstructed average QPU runtime latency for the single track hit selection QUBO. 
        The metrics are normalized by 1128 photon candidate events under a parameter set of 
        $S=1, R=200$, and $T_a=20.0\,\mu\text{s}$. All values are presented in milliseconds.
    }
    \label{fig:qpu_timing_part_a}
\end{figure}

\subsection{Performance of Simultaneous Track Candidate Association}

The second QUBO formulation was evaluated using cluster candidates obtained
after the detector hit filtering and clustering stage.  The purpose of this
test is not to refit the final track parameters, but to examine whether the
quantum annealer can identify geometrically consistent cluster triplets and
select several local track candidates within the same QPU submission.

The DAMSA detector geometry used in this study is optimized to minimize the amount of pile-up entering the tracking section. After the hit selection stage,
the number of additional track candidates beyond the two signal photon
showers is therefore small.  The association problem studied here should
therefore be interpreted as a proof of principle demonstration of simultaneous
QPU based cluster association in a low pileup detector environment, rather
than as a benchmark for a generic high occupancy tracking detector.

\subsubsection{Evaluation Definitions}

For this study, each reference track candidate is defined by the cluster
identifier assigned by the detector hit clustering stage in the simulation
based benchmark sample.  If a reference candidate appears in $L$ detector
layers, the set of reference nodes is written as
\begin{equation}
T =
\{(l_1,c),(l_2,c),\ldots,(l_L,c)\},
\end{equation}
where $l_i$ denotes the detector layer and $c$ is the cluster identifier.

The QPU output consists of selected cluster triplets.  A selected triplet is
counted as correct when all three selected nodes belong to the same reference
track candidate.  For a reference candidate with length $L$, the number of
possible triplets is
\begin{equation}
N_{\mathrm{triplet}}(L)
=
{L \choose 3}.
\end{equation}
The triplet recovery fraction is then defined as
\begin{equation}
\epsilon_{\mathrm{triplet}}
=
\frac{
N_{\mathrm{correct~triplet}}
}{
{L \choose 3}
}.
\end{equation}
This quantity evaluates the local association quality directly at the level
where the QUBO is constructed.

After the QPU selected cluster triplets are obtained, they are assembled into
longer track candidates using the graph overlap rule described in the
previous section.  For each reference candidate, the matched layer fraction is
defined as
\begin{equation}
\epsilon_{\mathrm{layer}}
=
\frac{
N_{\mathrm{matched~layers}}
}{
N_{\mathrm{reference~layers}}
}.
\end{equation}
This quantity measures how much of the original reference candidate is
recovered after graph assembly.

\subsubsection{Cluster Triplet Recovery}

\begin{figure}[htbp]
    \centering
    \includegraphics[width=\linewidth]{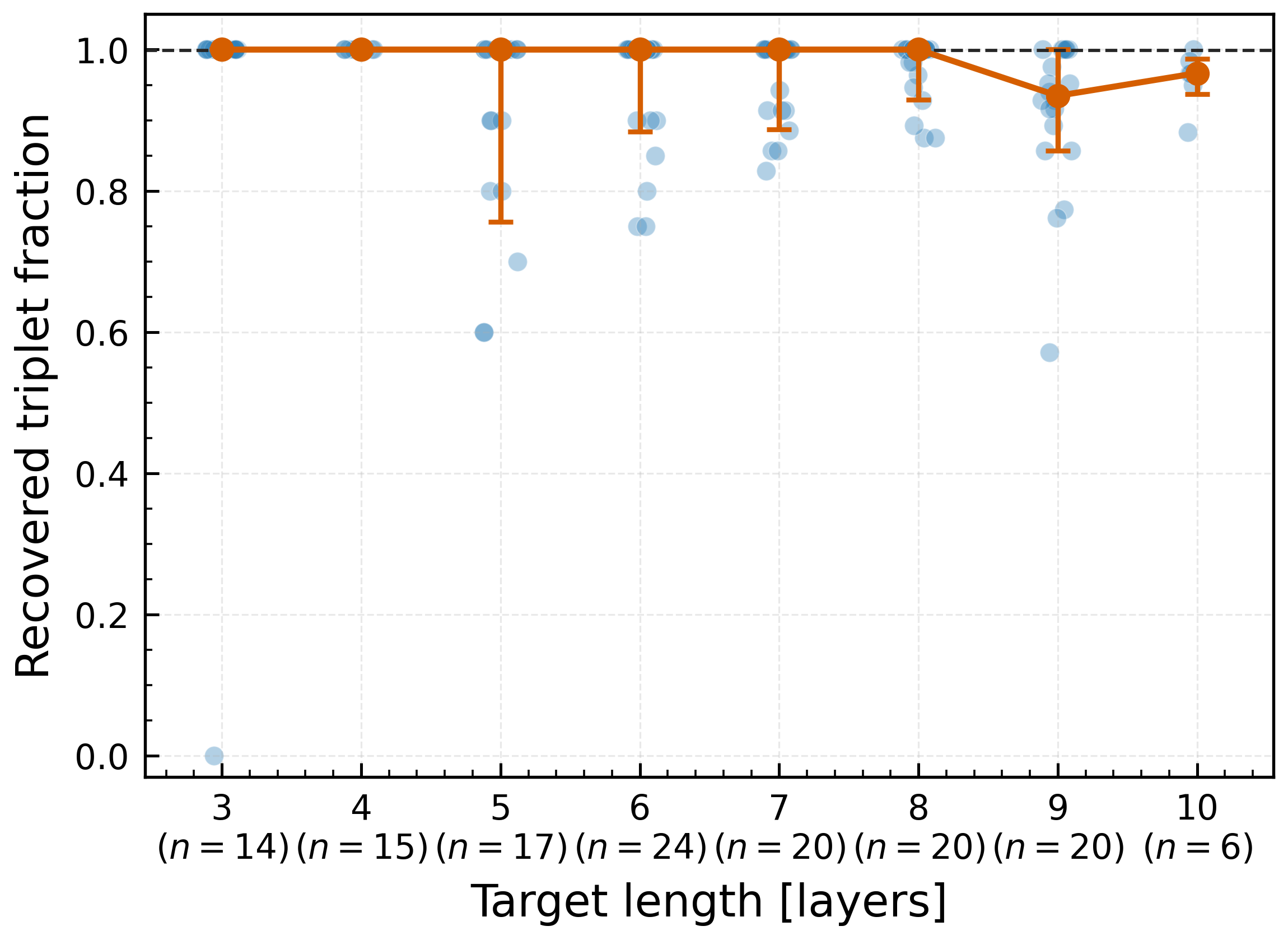}
    \caption{
        Recovery fraction of QPU selected cluster triplets.
        Each point corresponds to one reference track candidate. The orange
        markers show the median recovery fraction for each reference candidate
        length, with the vertical bars indicating the interval from 16\% to
        84\%. The number of reference candidates in each length bin is shown
        in the corresponding axis label.
    }
    \label{fig:qubo_triplet_recovery}
\end{figure}

Figure~\ref{fig:qubo_triplet_recovery} shows the recovery fraction of the
QPU selected cluster triplets as a function of the reference candidate
length.  The recovery fraction is evaluated using all possible three layer
subsets of each reference candidate.  A value of unity means that every
possible triplet belonging to the reference candidate was recovered by the selected block solutions.

In the evaluated sample, 136 reference candidates with at least three detector
layers are used for this evaluation. The median recovery remains close to
unity over most reference candidate lengths, showing that the QUBO formulation
usually identifies the correct cluster triplets. The spread observed in some
length bins reflects cases where a subset of the valid local triplets was not
selected in the sampled QPU solutions.  Since longer reference candidates
contain a larger number of possible triplets, they also provide more
opportunities for individual local associations to be missed.

\subsubsection{Track Candidate Recovery from QPU Triplets}

\begin{figure}[htbp]
    \centering
    \includegraphics[width=\linewidth]{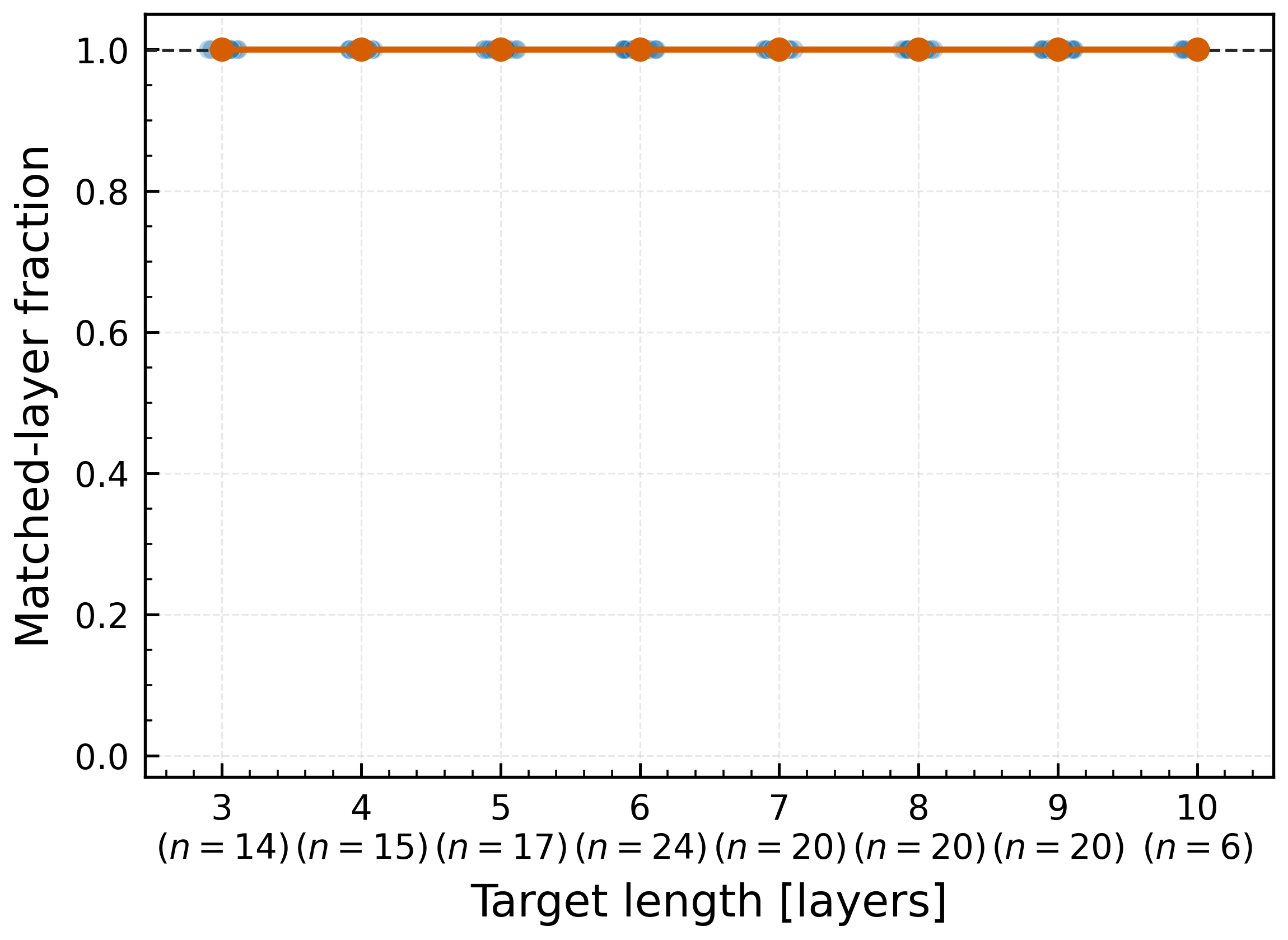}
    \caption{
        Matched layer fraction after graph assembly of QUBO selected cluster
        triplets. The selected triplets are connected when they share at least
        two identical layer-cluster nodes. Each connected component is
        interpreted as one reconstructed track candidate.
    }
    \label{fig:qubo_graph_chain_recovery}
\end{figure}

Figure~\ref{fig:qubo_graph_chain_recovery} shows the matched layer fraction
after the graph assembly step. Compared with the triplet recovery metric,
this observable evaluates the longer track candidates obtained from the
overlap graph.

The graph assembly recovers the reference track candidates with high
efficiency. This behavior is expected because even if a subset of local
cluster triplets is not selected, the remaining overlapping triplets can
still connect the same reference nodes into a single graph component. Thus,
the graph step provides a simple way to convert redundant QPU selected local
triplets into longer track candidates without introducing an additional line
fit, Kalman update, or residual based optimization. The final assembly is
based only on the discrete node consistency of the QPU selected triplets.

\subsubsection{QPU Resource Scaling for Simultaneous Track Candidate Association}

\begin{figure}[htbp]
    \centering
    \includegraphics[width=\linewidth]{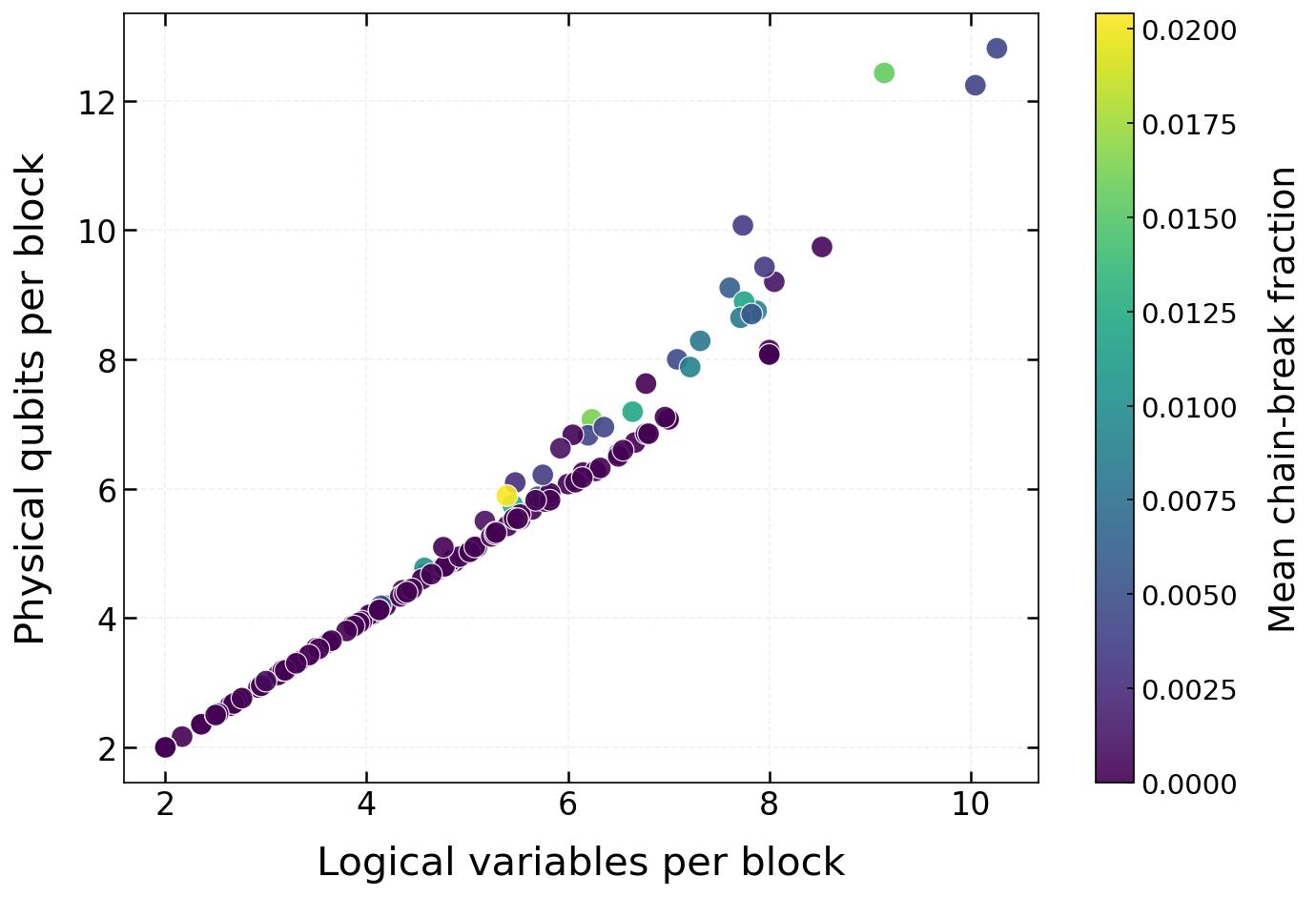}
    \caption{
        QPU resource usage for the simultaneous track candidate association
        QUBO.  The logical variable count and physical qubit usage are
        normalized by the number of three layer blocks included in each QPU
        submission.
    }
    \label{fig:qubo_cluster_resource_scaling}
\end{figure}

Figure~\ref{fig:qubo_cluster_resource_scaling} summarizes the QPU embedding
resources used by the cluster association QPU submissions.  Since multiple
independent three layer blocks are submitted together in one block diagonal
QUBO matrix, both the logical variable count and the physical qubit usage are
normalized by the number of blocks in the corresponding QPU call.

The observed scaling shows that the physical qubit cost per block increases
smoothly with the logical variable count per block.  This indicates that the
block matrix submission strategy keeps the embedding cost controlled while
reducing the number of separate QPU calls.  In the present low pileup sample,
the logical QUBO size remains modest because only a small number of cluster
candidates are present in each detector layer.  This resource study therefore
demonstrates the feasibility of the formulation on the current QPU, while
larger occupancy environments would require a dedicated scaling study.
\subsubsection{QPU Timing Model and Latency Deconstruction}

To evaluate the scalability and operational latency of the QPU during the simultaneous track candidate association phase, we apply the identical latency deconstruction model of Eqs. (59), (60), and (61) to establish a standardized benchmark baseline. This approach isolates the timing performance of the combinatorial association step from the initial hit-filtering stages.

For this multi-track problem-level evaluation, the hardware parameters were configured with an annealing time of $T_{\mathrm{a}} = 50.0\,\mu\mathrm{s}$, $R = 500$ reads, and $S = 4$ spin reversal transforms to ensure high-fidelity ground state searches within the larger, coupled QUBO space. Under this setup, the physical programming and embedding overhead per association task is measured as $T_{\mathrm{programming}} = 284.1\,\mathrm{ms}$. This robust configuration yields an average quantum annealing and sampling latency of $T_{\mathrm{sampling}} = 777.7\,\mathrm{ms}$, converging to a total mean end-to-end latency of $1137.5\,\mathrm{ms}$ per event, as illustrated in Fig. 9.

\begin{figure}[htbp]
    \centering
    \includegraphics[width=0.75\linewidth]{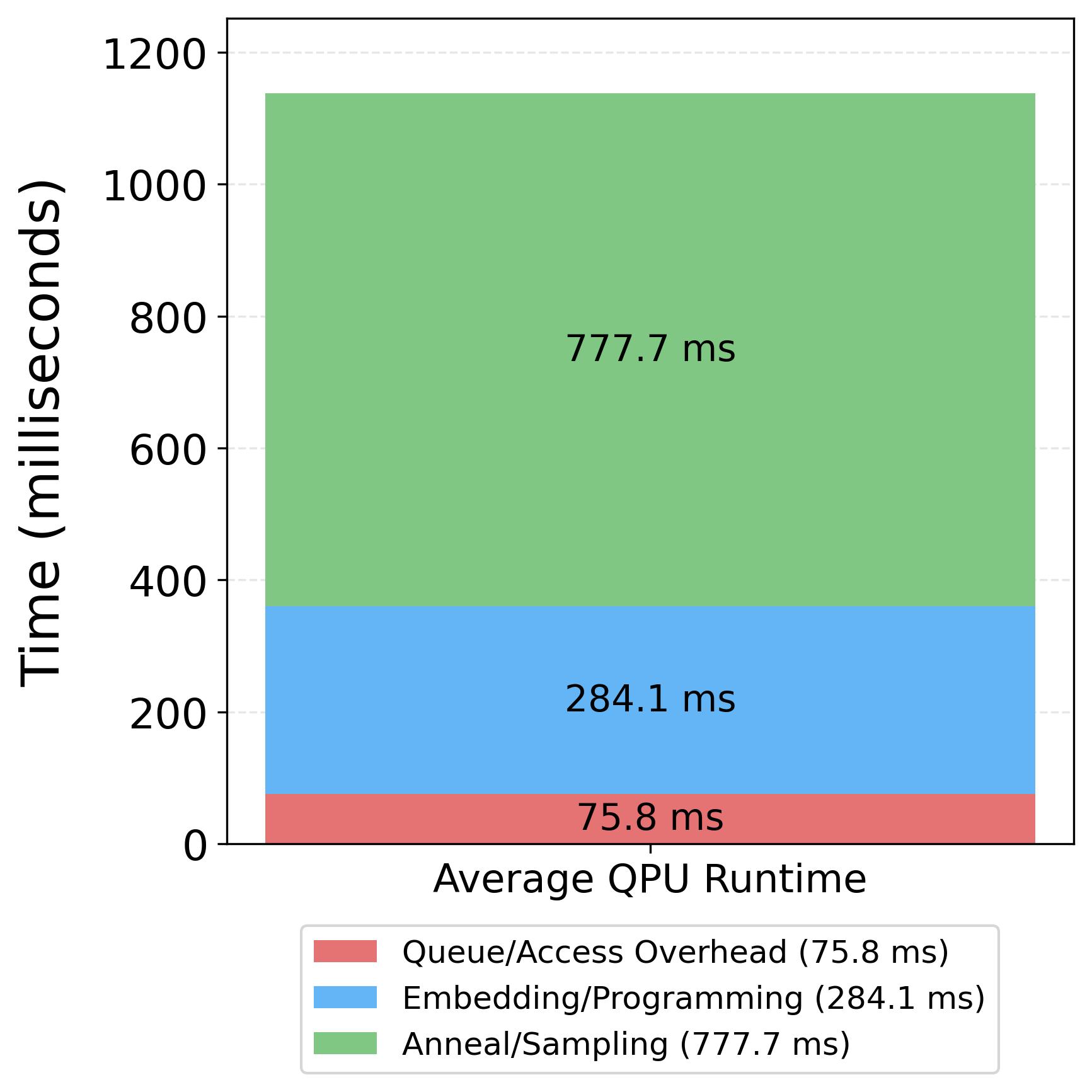}
    \caption{
        Deconstructed average QPU runtime latency for the simultaneous track candidate association QUBO. 
        The metrics are normalized by 75 target problem events under a parameter set of 
        $S=4, R=500$, and $T_a=50.0\,\mu\text{s}$. All values are presented in milliseconds.
    }
    \label{fig:qpu_timing_part_b}
\end{figure}

\section{Conclusions}

This work investigated two QUBO formulations for detector reconstruction in
the DAMSA detector environment.  The first formulation was designed for single
track hit selection inside a localized photon cluster region, where the goal
is to select a geometrically consistent hit in each detector layer while
rejecting ghost hit candidates.  The second formulation was designed for
simultaneous track candidate association, where cluster candidates from
different detector layers are associated into three dimensional track
candidates using QPU selected cluster triplets.

For the single track hit selection task, the QUBO method was compared with a
Kalman filter reconstruction using the same single photon sample.  The QPU
reconstruction produced position and angular residual distributions close to
those obtained with the Kalman method.  The Kalman filter gave slightly better
numerical resolutions, as expected from its continuous state update
procedure, but the QUBO method reproduced the overall residual shapes well.
The reduced $\chi^2$ values below unity and the small KL divergences indicate
that the QPU based discrete hit selection procedure achieves reconstruction
performance comparable to the conventional baseline in this detector
configuration.

For the simultaneous association task, all possible three layer combinations
among the ten detector layers were considered as QUBO blocks.  Each block can
select multiple valid cluster triplets, and several independent blocks are
submitted together in one block diagonal QUBO matrix.  This structure allows
one QPU submission to search for several local associations at the same time.
The selected cluster triplets are then assembled using a graph overlap rule.
This construction separates the quantum annealing part of the algorithm from
the graph assembly step: the QPU selects the local cluster triplets, while the
graph step connects triplets that share identical layer-cluster nodes.

The QPU resource studies show that the block matrix strategy provides a
practical way to submit many independent local QUBO problems in a reduced
number of QPU calls.  By normalizing the logical variable and physical qubit
counts by the number of submitted blocks, the embedding cost per local
reconstruction unit was shown to remain controlled for the low pileup events
studied here.

The present study should be viewed as a proof of principle demonstration.
The DAMSA detector geometry is optimized to reduce background activity near
the tracking section, and the selected association benchmark is therefore
relatively sparse compared with a generic high occupancy tracking
environment.  After the hit selection stage, the events are dominated by the
two energetic photons from ALP decay, while additional background induced
track candidates are rare.  Future studies should therefore test the same
formulation under larger cluster multiplicities, stronger pileup conditions,
and more realistic detector noise scenarios.  Further optimization of the
coefficients in QUBO energy terms, embedding strategy, and hybrid classical quantum postprocessing
may also improve the robustness of the method for more complex tracking
topologies.

Overall, the results demonstrate that QUBO formulations can be used both for
fine detector hit selection and for simultaneous track candidate association
in a particle detector reconstruction problem.  These results support further
studies of quantum annealer based reconstruction as a hybrid quantum classical tool for
detector environments where discrete hit association, ghost hit rejection, and
track candidate reconstruction dominate the combinatorial complexity.

\acknowledgments
The authors gratefully acknowledge the J\"ulich Supercomputing Centre for funding this project by providing computing time on the D-Wave Advantage\texttrademark{} System JUPSI through the J\"ulich UNified Infrastructure for Quantum computing (JUNIQ). This research was supported by 'Creation of the quantum information science R\&D ecosystem(based on human resources)' through the National Research Foundation of Korea(NRF) funded by the Korean government (RS-2023-NR057243). This research was supported by the National Research Foundation (NRF) funded by the Korean government (MSIT)(No. RS-2025-02221606 and No. RS-2026-25493297). 
\bibliography{ref}
\end{document}